\documentclass[11pt,letterpaper]{article}
\usepackage{color,graphicx,natbib,lscape,here,geometry,setspace,multirow,enumitem}
\usepackage[dvipsnames]{xcolor}
\usepackage{graphicx}
\usepackage{authblk}
\usepackage{silence}
\usepackage{rotating}
\usepackage{multirow}
\usepackage{booktabs}
\usepackage{color}
\usepackage{setspace}
\usepackage{algorithm2e}
\usepackage{enumitem}
\usepackage{tikz}

\usepackage{charter}
\usepackage[bitstream-charter]{mathdesign}

\usepackage[colorlinks=true,urlcolor=blue,linkcolor=blue,citecolor=blue]{hyperref}
\usepackage{bigstrut}

\usepackage{tabularx}
\usepackage{threeparttable}
\usepackage{dcolumn}

\usepackage{pdflscape}
\usepackage{afterpage}

\usepackage{longtable}
\usepackage{array}
\newcolumntype{C}[1]{>{\centering\arraybackslash}p{#1}}

\usepackage{bm}
\usepackage{comment}
\usepackage{mathtools}

\usepackage[hang,flushmargin]{footmisc} 
\usepackage[yyyymmdd]{datetime}
\usepackage[british]{babel}

\pdfoutput=1
 \usepackage{float}
 \restylefloat{table}

\usepackage[title,titletoc]{appendix}
\makeatletter

\renewenvironment{appendices}{%
    \begin{oldappendices}%
    \renewcommand{\thefigure}{\ifnum \c@section>\z@ \thesection.\fi\@arabic\c@figure}%
    \@addtoreset{figure}{section}%
    \renewcommand{\thetable}{\ifnum \c@section>\z@ \thesection.\fi\@arabic\c@table}%
    \@addtoreset{table}{section}%
}{%
    \end{oldappendices}%
}\makeatother

\usepackage{natbib}
\let\natbibcitet\citet
\renewcommand\citet{\bibpunct{(}{)}{,}{a}{,}{,}\natbibcitet}
\let\natbibcitep\citep
\renewcommand\citep{\bibpunct{(}{)}{;}{a}{,}{;}\natbibcitep}
\newcommand{\bi}{\begin{itemize}}
\newcommand{\ei}{\end{itemize}}
\newcommand{\be}{\begin{equation}}
\newcommand{\ee}{\end{equation}}
\defcitealias{ieo14}{IEO, 2014}

\long\def\symbolfootnote[#1]#2{\begingroup%
\def\thefootnote{\fnsymbol{footnote}}\footnote[#1]{#2}\endgroup}

\usepackage{sectsty}
\allsectionsfont{\normalsize}

\makeatletter
\def\ubar#1{\underline{\sbox\tw@{$#1$}\dp\tw@\z@\box\tw@}}
\def\obar#1{\overline{\sbox\tw@{$#1$}\dp\tw@\z@\box\tw@}}
\makeatother

\usepackage{bm}
\usepackage{caption}
\usepackage{subcaption}

\makeatletter\let\p@subfigure\thefigure\makeatother

\floatstyle{plaintop}
\restylefloat{table}

\usepackage{cleveref}
\crefname{chapter}{Chapter}{Chapters}
\crefname{section}{Section}{Sections}
\crefname{subsection}{Section}{Sections}
\crefname{subsubsection}{Section}{Sections}
\crefname{figure}{Figure}{Figures}
\crefname{table}{Table}{Tables}
\crefname{equation}{Equation}{Equations}
\crefname{appendix}{Appendix}{Appendices}
\crefname{appendices}{Appendix}{Appendices}

\crefname{appsec}{Appendix}{Appendices}

\usepackage{catoptions}
\makeatletter

\def\Autoref#1{%
  \begingroup
  \edef\reserved@a{\cpttrimspaces{#1}}%
  \ifcsndefTF{r@#1}{%
    \xaftercsname{\expandafter\testreftype\@fourthoffive}
      {r@\reserved@a}.\\{#1}%
  }{%
    \ref{#1}%
  }%
  \endgroup
}
\def\testreftype#1.#2\\#3{%
  \ifcsndefTF{#1autorefname}{%
    \def\reserved@a##1##2\@nil{%
      \uppercase{\def\ref@name{##1}}%
      \csn@edef{#1autorefname}{\ref@name##2}%
      \autoref{#3}%
    }%
    \reserved@a#1\@nil
  }{%
    \autoref{#3}%
  }%
}
\makeatother

\newcolumntype{d}[1]{D{.}{.}{#1}}

\title{Flexible shrinkage in high-dimensional Bayesian spatial autoregressive models}
\author[1]{Michael Pfarrhofer\thanks{Corresponding author: Michael Pfarrhofer, Department of Economics, WU Vienna University of Economics and Business. Address: Welthandelsplatz 1, 1020 Vienna, Austria. Email: \href{mailto:michael.pfarrhofer@wu.ac.at}{michael.pfarrhofer@wu.ac.at}. The authors would like to thank Maximilian B\"{o}ck, Niko Hauzenberger, Florian Huber and Gregor Zens for valuable comments and suggestions.}}
\author[2]{Philipp Piribauer}

\affil[1]{%
  \small{WU Vienna University of Economics and Business}}
\affil[2]{%
  \small{Austrian Institute of Economic Research (WIFO)}}
\date{\normalsize\today}


\def\equationautorefname~#1\null{%
  Eq.~(#1)\null
}
\def\equationautorefname~#1\null{
Eq.~(#1)\null
}

\usepackage[utf8, latin1]{inputenc}                 

\begin{document}
\maketitle\thispagestyle{empty}\normalsize

\begin{abstract}
\noindent This article introduces two absolutely continuous global-local shrinkage priors to enable stochastic variable selection in the context of high-dimensional matrix exponential spatial specifications. Existing approaches as a means to dealing with overparameterization problems in spatial autoregressive specifications typically rely on computationally demanding Bayesian model-averaging techniques. The proposed shrinkage priors can be implemented using Markov chain Monte Carlo methods in a flexible and efficient way. A simulation study is conducted to evaluate the performance of each of the shrinkage priors. Results suggest that they perform particularly well in high-dimensional environments, especially when the number of parameters to estimate exceeds the number of observations. For an empirical illustration we use pan-European regional economic growth data.
\end{abstract}

\bigskip
\begin{tabular}{p{0.2\hsize}p{0.65\hsize}} 
\textbf{Keywords:}  &Matrix exponential spatial specification, model selection, shrinkage priors, hierarchical modeling\\
\end{tabular}

\smallskip
\begin{tabular}{p{0.2\hsize}p{0.4\hsize}}
\textbf{JEL Codes:} &C11, C21, C52 \\
\end{tabular}
\vspace{0.6cm}

\bigskip
\newpage
\section{Introduction}
\label{sec:introduction}
In the regional economic literature, spatial econometric model specifications have gained momentum in empirical research as a means to explicitly account for spillover effects between geographically structured units. Increasing availability of data often results in (spatial autoregressive) models where the number of observations is relatively small compared to the number of potential covariates. Standard estimation techniques in such environments therefore typically result in imprecise parameter estimates. In the case of severe overparameterization, where the dimensionality of the regressors exceeds the number of observations, standard estimation approaches may even be infeasible.

Bayesian model averaging techniques to alleviate the problems of overparameterization in spatial autoregressive models have been proposed \citep{GEAN:GEAN703} and widely applied, particularly in the empirical study of regional economic growth (see, for example, \citealt{doi:10.1080/17421770802353758}, \citealt{JAE:JAE2277}, \citealt{doi:10.1080/00343404.2012.678824}, \citealt{GEAN:GEAN12057}, or \citealt{cuaresma2018human}). These approaches use weighted averages of parameter estimates based on a multitude of potential combinations of explanatory variables, rather than relying on inference based on a single model specification (for extensive discussions see \citealt{steel2017bma} or \citealt{koop2003}). Bayesian model averaging, however, involves the computation of marginal likelihoods for integrating out the underlying model uncertainty. In contrast to classical linear model frameworks, no closed-form solutions for marginal likelihoods in spatial autoregressive specifications are available \citep{GEAN:GEAN703}, resulting in a severe computational burden especially in high-dimensional estimation problems.

Recent advances in the spatial econometric literature focus on extensions and generalizations of standard spatial autoregressive specifications. For example, some extensions aim at explicitly controlling for unobserved heterogeneity by finite mixture or threshold specifications (\citealt{Piribauer2016}; \citealt{CORNWALL2017148}), allowing for heterogeneous parameters across space (\citealt{GEAN:GEAN12152}), heteroskedastic specifications of the innovations (\citealt{doi:10.1177/016001769702000107}), considering continuous spatial effects (\citealt{Laurini2017}), or accounting for uncertainty in the spatial weight matrix specification and the underlying nature of spillover processes (\citealt{LESAGE2007190}; \citealt{JORS:JORS12188}; \citealt{GEAN:GEAN12057}), among several others. However, such extensions to the spatial autoregressive modeling framework further increase the emanating computational burden of Bayesian model averaging, rendering the approach computationally intractable.

For spatial autoregressive model specifications, work by \cite{Piribauer2016} and \cite{doi:10.1080/17421772.2016.1227468}, for example, uses Bayesian stochastic search variable selection priors \citep[SSVS,][]{doi:10.1080/01621459.1993.10476353} as an alternative to Bayesian model averaging. The comparative flexibility of this approach allows to easily extend and generalize the basic framework to more complex specifications.\footnote{The computational flexibility of these shrinkage priors comes from the fact that they can easily be implemented in standard Bayesian Markov-chain Monte Carlo (MCMC) algorithms without the need for calculating marginal likelihoods \citep{doi:10.1080/01621459.1993.10476353}.} Specifically, the proposed hierarchical modeling approach assumes coefficients of a saturated model under scrutiny to come from a mixture of two Gaussians centered on zero with a spike and a slab component (low and high variance). A binary latent indicator thereby identifies promising subsets of predictors by shrinking unimportant components towards zero. Despite its elegance and appealing theoretical properties, difficulties arise for stochastic search variable selection in large data sets due to the necessity of stochastic search over an enormous space. This implies slow mixing and convergence during estimation \citep{doi:10.1080/01621459.2014.960967}. Flexible alternatives for shrinkage in high-dimensional econometric frameworks have therefore been advocated \citep{polson2010,griffin2017}.

In this paper we aim at generalizing variants of the Normal-Gamma \citep[NG,][]{griffin2010} and the Dirichlet-Laplace \citep[DL,][]{doi:10.1080/01621459.2014.960967} shrinkage priors to spatial autoregressive specifications. For the purpose of illustrating prior specific properties in the presence of spatial dependence, we carry out an extensive simulation study using synthetic data sets, considering different scenarios of the number of available covariates and degrees of sparsity of the coefficient vector. The paper moreover considers matrix exponential spatial specifications (MESS), introduced by \cite{LESAGE2007190} to model global spillover processes. Our results indicate that conventional stochastic search variable selection priors in the spirit of \cite{doi:10.1080/01621459.1993.10476353} work well in relatively low-dimensional settings, where the number of potential covariates do not exceed those of the observations. In high-dimensional modeling frameworks, however, both the Normal-Gamma and Dirichlet-Laplace shrinkage prior exhibit stellar empirical properties. They provide a high degree of adaptiveness of shrinkage, with the Normal-Gamma excelling in terms of precision, while the Dirichlet-Laplace performs particularly well in terms of point estimates.

The remainder of this paper is organized as follows. \Autoref{sec:econometrics} presents the Bayesian spatial econometric framework, followed by the introduction of Normal-Gamma and Dirichlet-Laplace shrinkage priors to spatial autoregressive specifications in \Autoref{sec:shrinkage}. A simulation study as a means to comparing the properties of the proposed shrinkage priors is presented in \Autoref{sec:simulation}. \Autoref{sec:application} illustrates the performance of the proposed shrinkage priors using pan-European regional economic growth data. \Autoref{sec:conclusions} concludes.

\section{Econometric framework}
\label{sec:econometrics}
We start by considering a model of the form
\begin{equation}
\bm{S}(\bullet)\bm{y} = \bm{X}\bm{\beta} + \bm{\epsilon}, \quad \bm{\epsilon}\sim\text{N}(\bm{0}, \bm{\Omega}),
\label{eq:messdm}
\end{equation}
where $\bm{y}$ is an $N$-dimensional vector of dependent variables and $\bm{S}(\bullet)$ describes a linear transformation dependent on a not yet specified parameter. $\bm{X}$ is an $N\times K$ matrix of explanatory variables (with a vector of ones in the first column), and $\bm{\beta} = (\beta_1, \hdots, \beta_K)'$ is a $K$-dimensional vector of parameters. We assume the error term $\bm{\epsilon}$ to be normally distributed with zero mean and $N\times N$ variance-covariance matrix $\bm{\Omega}$. 

Standard econometric models capturing spatial spillover effects are thoroughly discussed in \citet{lesage-pace_introductionspatial}. Conventional spatial autoregressive models (SAR) typically set $\bm{S}(\xi) = (\bm{I}_N - \xi \bm{W})$, where $\bm{W}$ is an $N\times N$ exogenous right stochastic spatial weights matrix. If observations $i=1,\dots, N$ and $j\neq i$ are considered neighbors, then $W_{ij} \neq 0$, otherwise $W_{ij} = 0$. By convention $W_{ii} = 0$, meaning that a region or other spatial unit is not considered a neighbor to itself \citep{lesage-pace_introductionspatial}. $\xi$ is a (scalar) spatial parameter with stability condition $|\xi|<1$. The inverse of $\bm{S}(\xi)$ under the specifying assumptions for $\bm{W}$ and $\xi$ can be expressed as $(\bm{I}_N-\xi\bm{W})^{-1}= \sum_{l=0}^{\infty} \xi^l \bm{W}^l$, implying global geometric decay over space. An alternative to such a geometric decay pattern is given by the matrix exponential spatial specification (MESS), as proposed by \cite{LESAGE2007190}, where we set 
\begin{equation}
\bm{S}(\rho) = \exp(\rho \bm{W}) = \sum_{l=0}^{\infty} \rho^l \bm{W}^l (l!)^{-1},\label{eq:matrixexponential}
\end{equation}
with the (scalar) parameter $\rho$ taking the role of measuring spatial dependence. Consequently, these modeling approaches nest the classical linear regression model in the case where the respective spatial dependence parameter is equal to zero. The main difference between MESS and SAR models is that spatial externalities are modeled by an exponential rather than a geometric decay.\footnote{It is moreover worth noting that a simple model extension frequently used in empirical applications also incorporates an explicit spatial structure in the exogenous variables by including a spatial lag of $\bm{X}$ in \autoref{eq:messdm}. Such a specification is commonly referred to as a spatial Durbin model specification \citep{lesage-pace_introductionspatial}.}

MESS is advantageous to the conventional SAR approach especially in a Bayesian framework. This is due to the properties of matrix exponentials presented in \citet{LESAGE2007190},
\begin{enumerate}[label=(\roman*),wide=0pt,leftmargin=*]
  \item $\bm{S}(\rho)$ is non-singular, 
  \item $\bm{S}(\rho)^{-1} = \exp(\rho \bm{W})^{-1} = \exp(-\rho \bm{W})$, and 
  \item $\det(\exp(\rho \bm{W})) = \exp(\text{tr}(\rho \bm{W}))$,
\end{enumerate}
which facilitate the likelihood function given in \autoref{eq:likelihood}. Albeit both specifications typically produce rather similar estimates and inference (see, for example, \citealt{LESAGE2007190}, \citealt{GEAN:GEAN12057}, or \citealt{STRAU2017221}), matrix exponential spatial specifications have some potential computational advantages, specifically in high-dimensional data environments. A correspondence between spatial coefficients in the conventional SAR and MESS is stated by \citet{LESAGE2007190}, where $\xi \approx 1 - \exp(\rho)$. However, it is worth noting that \citet{DEBARSY20151} stress that no precise one-to-one correspondence can be established and the two spatial specifications should not be considered perfect substitutes, due to $\rho \in (-\infty,\infty)$ while stability conditions in the conventional SAR case require the parameter space of $\xi$ to be constrained \citep{lesage-pace_introductionspatial}. Assuming a normally distributed error term as in \autoref{eq:messdm}, the likelihood of the model is given by
\begin{equation}
p(\bm{y}|\bm{X},\bm{\beta},\sigma^2,\rho) = (2\pi)^{-N/2} \det(\bm{\Omega})^{-1/2} \exp\left(-\bm{\varepsilon}'\bm{\Omega}^{-1}\bm{\varepsilon} / 2\right),\label{eq:likelihood}
\end{equation}
where we define $\bm{\varepsilon} = \left(\bm{S}(\rho)\bm{y} - \bm{X}\bm{\beta}\right)$ for notational convenience.

\section{Flexible shrinkage in high-dimensional Bayesian spatial autoregressive models}\label{sec:shrinkage}
One conventional approach for cases where the vector of coefficients is expected to be sparse, but without prior knowledge which variables to exclude, is given by penalized least squares. A prominent example is the least absolute shrinkage and selection operator (LASSO) introduced by \citet[][]{10.2307/2346178}. This paper stresses the correspondence between the conventional LASSO and a Bayesian approach involving independent double-exponential priors on regression coefficients, a notion further elaborated on in \citet{doi:10.1198/016214508000000337}. Given the choice of a specific mixture distribution $\text{D}$, following \citet{griffin2010}, these priors can typically be expressed as
\begin{equation}
\beta_r|\Psi_r \sim \text{N}(0,\Psi_r), \quad \Psi_r \sim \text{D}.
\end{equation}
for $r=1,\dots,K$. In this setting, the marginal distribution for $\beta_r$ has heavier than normal tails but places substantial mass on zero. \citet{griffin2010} indicate that the standard discrete spike-and-slab prior (where a variant is presented in \cref{subsec:ssvs}) may be represented in this form, dependent on the mixture distribution being chosen accordingly. 

In addition, various other cases such as the double-exponential mentioned above (leading to the Bayesian LASSO, which is closely tied to the Normal-Gamma prior presented later on), arise for different mixing distributions \citep{doi:10.1198/016214508000000337}. Note that in this basic framework, the specific shape of $\text{D}$ depends on the deterministic choice of prior hyperparameters. More flexibility and adaptive shrinkage can be achieved by introducing additional hierarchical layers of priors at this stage. Approaches in this spirit are termed global-local shrinkage priors \citep{polson2010}, and seem to perform well in high-dimensional settings \citep[for instance in time-series analysis, see][]{bittosfs,kastner,doi:10.1080/07350015.2016.1256217,feldkircherkastnerhuber}. In the following, we discuss two alternatives of continuous global-local shrinkage, the Normal-Gamma and the Dirichlet-Laplace shrinkage prior.

\subsection{Normal--Gamma shrinkage prior}\label{subsec:ng}
A variant of the Normal-Gamma global-local shrinkage prior, as proposed by \cite{griffin2010}, is given by a scale mixture of Gaussians,
\begin{align}
\beta_r | \psi_r &\sim \text{N}(0,2\lambda^{-2}\psi_r),\\ 
\psi_r &\sim \text{G}(\theta,\theta),\nonumber\\
\lambda^2 &\sim \text{G}(d_0,d_1),\nonumber
\end{align}
where $\psi_r$ is an idiosyncratic scaling parameter following a Gamma distribution that involves parameter specific shrinkage and can be collected in a vector $\bm{\psi} = (\psi_1,\hdots,\psi_K)'$. Overall shrinkage is governed by the global parameter $\lambda$ that also follows a Gamma distribution. According to \citet{polson2010}, this setup reflects the necessity of noise reduction by shrinking all coefficient means to zero based on the global parameter, while allowing for signals to override this effect using local scaling parameters. Hyperparameters must be set by the researcher, where standard approaches in the literature include $d_0 = d_1 = 0.01$, and $\theta = 0.1$ as default. This implies heavy overall shrinkage of the parameters stemming from the global component but provides enough flexibility to detect individual non-zero coefficients if necessary.\footnote{In addition, it is worth mentioning that $\theta$ may be also integrated out, as for instance described in \citet{doi:10.1080/07350015.2016.1256217}. This could easily be implemented within the given structure. However, for the sake of simplicity, we refrain from doing so.} Setting $\theta = 1$ would present the case leading to the Bayesian LASSO \citep{doi:10.1198/016214508000000337}. Deriving the posterior distribution of $\psi_r$, one finds that it follows a generalized inverse Gaussian distribution \citep{griffin2010},
\begin{equation}\label{eq:ng-psi}
p(\psi_r | \lambda, \beta_r) \sim \text{GIG}(\theta - 1/2, \beta_r^2,\theta \lambda^2),
\end{equation}
with $\text{GIG}(\zeta,\chi,\varrho)$ being parameterized such that its density $f(x) \propto x^{\zeta-1}\exp\{-(\chi/x + \varrho x)/2\}$, while the conditional posterior distribution of the global parameter is a Gamma distribution with
\begin{equation}\label{eq:ng-lambda}
p(\lambda^2 | \bm{\psi}) \sim \text{G}\left(d_0 + \theta K, d_1 + 2^{-1} \theta \sum_{r=1}^K \psi_r\right).
\end{equation}
This setup defines the prior variance-covariance matrix on $\bm{\beta}$, denoted by $\ubar{\bm{\Sigma}}$, that is updated during estimation. We set the prior coefficient vector denoted by $\ubar{\bm{\beta}} = \bm{0}$. In the case of the NG prior, the variance-covariance matrix is given by $\text{diag}(\ubar{\bm{\Sigma}}) = \bm{\psi}$. The conditional posterior for coefficients $\bm{\beta}$ is of typical form \citep{koop2003},
\begin{align}\label{eq:theta-ssvs}
p(\bm{\beta}|\bullet) &\sim \text{N}\left(\obar{\bm{\beta}}, \obar{\bm{\Sigma}}\right),\\
\obar{\bm{\Sigma}} &= \left(\ubar{\bm{\Sigma}}^{-1} + \bm{X}'\bm{\Omega}^{-1}\bm{X}\right)^{-1},\nonumber\\
\obar{\bm{\beta}} &= \obar{\bm{\Sigma}}\left(\ubar{\bm{\Sigma}}^{-1}\ubar{\bm{\beta}} + \bm{X}'\bm{\Omega}^{-1}\bm{S}(\rho)\bm{y}\right).\nonumber
\end{align}

\subsection{Dirichlet--Laplace prior}\label{subsec:dl}
Even though working well empirically, \citet{doi:10.1080/01621459.2014.960967} note that many aspects of global-local shrinkage priors based on scale mixtures of Gaussians are poorly understood theoretically and stress the necessity of simultaneous consideration of marginal properties of shrinkage priors and the joint distribution of obtained parameters.

As an alternative, we continue with the DL prior suggested by \cite{doi:10.1080/01621459.2014.960967}. As opposed to the three hyperparameters to be set by the researcher in the NG case, a single hyperparameter suffices for establishing the DL setup. Similar to the NG prior, however, it is composed hierarchically of global and local shrinkage parameters, and may be structured as follows:
\begin{align}\label{eq:dl-start}
\beta_r &\sim \text{N}(0, \varphi_r \phi_r^2 \tau^2)\\
\varphi_r &\sim \text{Exp}(1/2)\nonumber\\
\phi &\sim \text{Dir}(a,\hdots,a)\nonumber\\
\tau &\sim \text{G}(Na, 1/2).\nonumber
\end{align}
The local parameters are denoted by $\varphi_r$ and assigned an exponential prior distribution. In contrast to the single global parameter $\lambda$ in the case of the NG prior, the DL approach uses a vector of scales $(\phi_1 \tau, \hdots, \phi_K \tau)$ to provide more flexibility regarding idiosyncratic coefficient shrinkage, where $\bm{\phi} = (\phi_1,\hdots,\phi_K)$ is defined to lie in the ($K-1$)-dimensional simplex $\mathcal{S}^{K-1} = \{\mathbf{x} = (x_1, \hdots, x_K)':\ x_r\geq 0,\ \sum_{r=1}^{K} x_r = 1\}$ and is assigned a Dirichlet prior distribution with hyperparameter $a$ controlling the tightness of the prior. This quantity may again be integrated out as shown in \citet{doi:10.1080/01621459.2014.960967}. Based on theoretical results obtained by \citet{doi:10.1080/01621459.2014.960967}, standard deterministic choices include $a = 1/K$ (the default setting used for the simulation study), but different choices such as $a = 1/2$ are justifiable theoretically. Notice that this allows the DL prior setup to be dependent on the dimensionality of the underlying model on theoretical grounds, different to both the SSVS and NG prior.

In the case of the DL prior, \citet{doi:10.1080/01621459.2014.960967} show that $\varphi_r$ can be sampled efficiently involving independent inverse Gaussian distributions. This is done by obtaining $\tilde{\varphi}_r|\phi,\bm{\beta}$ from a reparameterization of the generalized inverse Gaussian distribution with $\mu_r = \phi_r\tau/|\beta_r|$,
\begin{equation}\label{eq:dl-varphi}
p(\tilde{\varphi}_r|\phi,\bm{\beta}) \sim \text{GIG}\left(-1/2,1,\mu_r^{-2}\right),
\end{equation}
where we subsequently set $\varphi_r = 1/\tilde{\varphi}_r$ to obtain draws from the conditional posterior distribution of $\varphi_r$. The global shrinkage component $\tau$ is again sampled from a generalized inverse Gaussian distribution
\begin{equation}\label{eq:dl-tau}
p(\tau|\phi,\bm{\beta}) \sim \text{GIG}\left(1-K, 2\sum_{r=1}^{K}|\beta_r|/\phi_r,1\right).
\end{equation}
In order to sample $\bm{\phi}|\bm{\beta}$ we rely on Theorem 2.1 in \citet{doi:10.1080/01621459.2014.960967} and draw auxiliary variables $T_1,\hdots,T_K$ independently, with $T_r\sim\text{GIG}(a-1, 2|\beta_r|, 1)$. Afterwards, we set $\phi_r= T_r/\sum_{j=1}^{K} T_j$. \citet{doi:10.1080/01621459.2014.960967} indicate this step as an important feature of their setup, as it significantly accelerates mixing and convergence. Consequently, the structure of the prior variance-covariance matrix $\ubar{\bm{\Sigma}}$ is updated, where $\text{diag}(\ubar{\bm{\Sigma}}) = (\varphi_1 \phi_1^2 \tau^2, \hdots, \varphi_K \phi_K^2 \tau^2)'$. The coefficient vector is then sampled analogously to the NG prior, based on \autoref{eq:theta-ssvs}.

\subsection{Priors for remaining parameters}\label{subsec:priors}
Without loss of generality, we assume a homoskedastic error variance $\bm{\Omega} = \sigma^2 \bm{I}_N$, where $\bm{I}_N$ is an $N\times N$-dimensional identity matrix.\footnote{The generic notation in \autoref{eq:messdm}, however, allows for various specifications of the variance-covariance matrix $\bm{\Omega}$, for instance reflecting heteroskedastic error terms \citep[see, for instance,][in a spatial context]{doi:10.1177/016001769702000107}.} To complete the prior setup, we have to elicit prior distributions for $\sigma^2$ and the spatial dependence parameter $\rho$. Here we use standard specifications in the literature \citep{lesage-pace_introductionspatial}. Specifically, we impose an inverse Gamma prior on the variance of the error term, $\sigma^2 \sim \text{G}^{-1}(\ubar{a},\ubar{b})$ with scalar hyperparameters $\ubar{a}$ and $\ubar{b}$ that we set equal to $0.01$, rendering them rather uninformative. Conditioning on all other parameters of the model and the data we obtain a conditional posterior density for the error variances,
\begin{align}\label{eq:sigma2}
p(\sigma^2|\bullet) &\sim\text{G}^{-1}(\obar{a},\obar{b}),\\
\obar{a} &= \ubar{a} + N/2,\nonumber\\
\obar{b} &= \ubar{b} + \bm{\varepsilon}'\bm{\varepsilon}/2.\nonumber
\end{align}

The quantities for $\bm{\beta}$ and $\sigma^2$ are standard and can be sampled efficiently using Gibbs sampling (\citealt{koop2003}). For $\rho$, we follow \cite{LESAGE2007190} and \cite{lesage-pace_introductionspatial}, by eliciting a normal prior $\rho \sim \text{N}(0,c)$, where $c$ may be chosen dependent on the prior belief of the researcher regarding the strength of spatial association in the data. For the simulation study, we again use a rather uninformative specification and choose $c = 10$. Since the posterior distribution of $\rho$ conditional on all other quantities of the model is in general not of well-known form,
\begin{equation}
p(\rho|\bullet) \propto \exp\left(-\bm{\varepsilon}'\bm{\Omega}^{-1}\bm{\varepsilon}/2\right)p(\rho),\label{eq:rho}
\end{equation}
it is sampled by a Metropolis-within-Gibbs step. We follow the standard approach by proposing a new value from a Normal distribution, $\rho^{\ast}\sim\text{N}(\rho_{t-1},\varsigma)$, where the subscript $t-1$ denotes the value of the parameter from the previous iteration of the sampling algorithm and $\varsigma$ is a tuning parameter. The acceptance probability of the proposal is calculated using
\begin{equation}
\min\left[1,\frac{p(\rho^\ast|\bullet)}{p(\rho_{t-1}|\bullet)}\right].\nonumber
\end{equation}
If the proposal is accepted, we set $\rho_{t} = \rho^{\ast}$. Otherwise, the proposal is rejected, and we retain the value from the previous draw. The process of generating proposals for $\rho$ is tuned during half of the burn-in phase of the algorithm by incrementally increasing or decreasing the variance of the proposal distribution $\varsigma$ to yield an acceptance rate for the parameter between $0.2$ and $0.4$. An overview of the algorithm employed can be found in \cref{app:posteriors}.

\section{Simulation study}
\label{sec:simulation}
In this section, we evaluate the empirical properties and merits of the proposed shrinkage priors. Estimates are obtained using the Normal-Gamma (NG) and Dirichlet-Laplace (DL) shrinkage priors sketched above. As a benchmark specification, we compare the results of both setups with a stochastic search variable selection (SSVS) prior put forward by \cite{doi:10.1080/01621459.1993.10476353} and applied to spatial autoregressive models by \cite{Piribauer2016} and \cite{doi:10.1080/17421772.2016.1227468}. The SSVS prior as a means to introducing shrinkage on the slope coefficients mimics Bayesian model-averaging frameworks. Details on the SSVS setup along with the concurrent Markov-chain Monte Carlo (MCMC) sampling algorithm are given in Appendix \ref{subsec:ssvs}. To further illustrate and underline the necessity of applying shrinkage in cases where the coefficient vector is sparse, we also provide results for the case where a rather uninformative prior variance-covariance matrix $\ubar{\bm{\Sigma}} = 1000 \bm{I}_N$ is used. This approach represents the basic setup given in \citet{LESAGE2007190}, and is labeled \textit{None}, referring to the fact that no shrinkage is applied. For each specification employed, we use $2,000$ iterations, discarding the initial $1,000$ draws as burn-in. Inference is then based on the $T=1,000$ posterior draws for all parameters, where point estimates are calculated using the median of the respective sample.

\subsection{Data Generating Process}\label{sec:dgp-sim}
Simulating a synthetic data set for different data generating processes requires decisions on the number of observations $N$ and explanatory variables $K$. Moreover, we have to set parameter values for the coefficients $\tilde{\bm{\beta}}$, the variance of the error terms $\tilde{\sigma}^2$ and the spatial autoregressive parameter $\tilde{\rho}$. We consider varying degrees of sparsity with respect to the coefficient vector. The final data generating process, reflecting the basic modeling approach as in \autoref{eq:messdm}, has the form
\begin{equation}
\tilde{\bm{y}} = \exp(-\tilde{\rho} \tilde{\bm{W}})\left(\tilde{\bm{Z}}\tilde{\bm{\beta}} + \bm{\epsilon}\right), \quad \tilde{\bm{\epsilon}}\sim\text{N}(0, \tilde{\sigma}^2 \bm{I}_N),
\end{equation}
where the first column in $\bm{Z}$ contains an intercept. Robustness for different combinations of parameters and generated variables is achieved via repeating each of the basic simulations sketched below for $100$ times and relying on stochastically setting most of the quantities involved. In particular, we set the required parameters as follows:
\begin{enumerate}[label=(\roman*),wide=0pt,leftmargin=*]
  \item The values for each of the $k = 1,\hdots,K-1$ (non-constant) explanatory variables $\bm{z}_k = [z_{k1},\hdots,z_{kN}]'$ for each observation $i = 1,\hdots,N$ come from a standard normal distribution, $z_{ki}\sim\text{N}(0,1)$ and are collected in the matrix $\bm{Z} = (\bm{\iota}_N, \bm{z}_1,\hdots,\bm{z}_{K-1})$ where $\bm{\iota}_N$ is an $N$-dimensional vector of ones. For all simulations, we use a sample size of $N = 100$ and consider $K \in \{50, 100, 150, 200\}$.
  \item For the longitude and latitude coordinates of the observations we randomly generate points within a unit square and construct a row-stochastic matrix $\bm{W}$ based on a five-nearest neighborhood specification.\footnote{Along the lines of \citet{STRAU2017221} we do not use a truncated Taylor series expansion of the matrix exponential for calculations in this article, but an efficient implementation of this operation in R.} It is worth mentioning that the specification of the number of neighbors appeared to have no discernible effect on the results obtained during simulations.
  \item Sparsity of the coefficient vector is governed by setting a maximum number of non-zero parameters. We evaluate two different scenarios, where the number of $\tilde{\beta}_k \neq 0$ does not exceed $q \in \{10, 20\}$. As we also intend to test adaptiveness of shrinkage, half of the non-zero parameters are generated to be close to zero; the other half is allowed to exhibit greater variation, while we always include a non-zero effect on the intercept. For the slope coefficients corresponding to the non-constant covariates in $\bm{Z}$, this is achieved by simulating ($K-1$) \textit{true} coefficient values $\tilde{\beta}_{1:(q/2)} \sim \text{N}(0,1)$ and $\tilde{\beta}_{(q/2+1):q} \sim \text{N}(0,5)$, where the parameter for the intercept always comes from the latter distribution. The remaining coefficients are set to zero in a deterministic fashion, $\tilde{\beta}_{(q+1):K} = 0$.
  \item The homoskedastic variance of the error terms is deterministically set to $\sigma^2 = 1$.
  \item The spatial dependence parameter is simulated using a zero-mean Gaussian distribution, $\tilde{\rho} \sim \text{N}(0,3)$.
\end{enumerate}

\subsection{Diagnostics}\label{sec:diagnostics}
We briefly consider diagnostics regarding obtained draws from the posterior distributions by evaluating trace plots and estimated densities. This allows us to present both the obtained marginal posterior densities for the respective parameters, and to discuss mixing and convergence properties. Diagnostic plots of non-zero coefficients typically mirror the well-known picture resulting from standard applications without introducing shrinkage. For illustrative purposes, we thus pick examples with true coefficients being equal to zero. A low-dimensional problem is given by choosing $q = 10$ and $K=50$, while the case of $K=150$ serves as high-dimensional scenario where applying shrinkage is required to find meaningful results on an acceptable level of precision.
\begin{figure*}[t]
  \includegraphics[width=\textwidth]{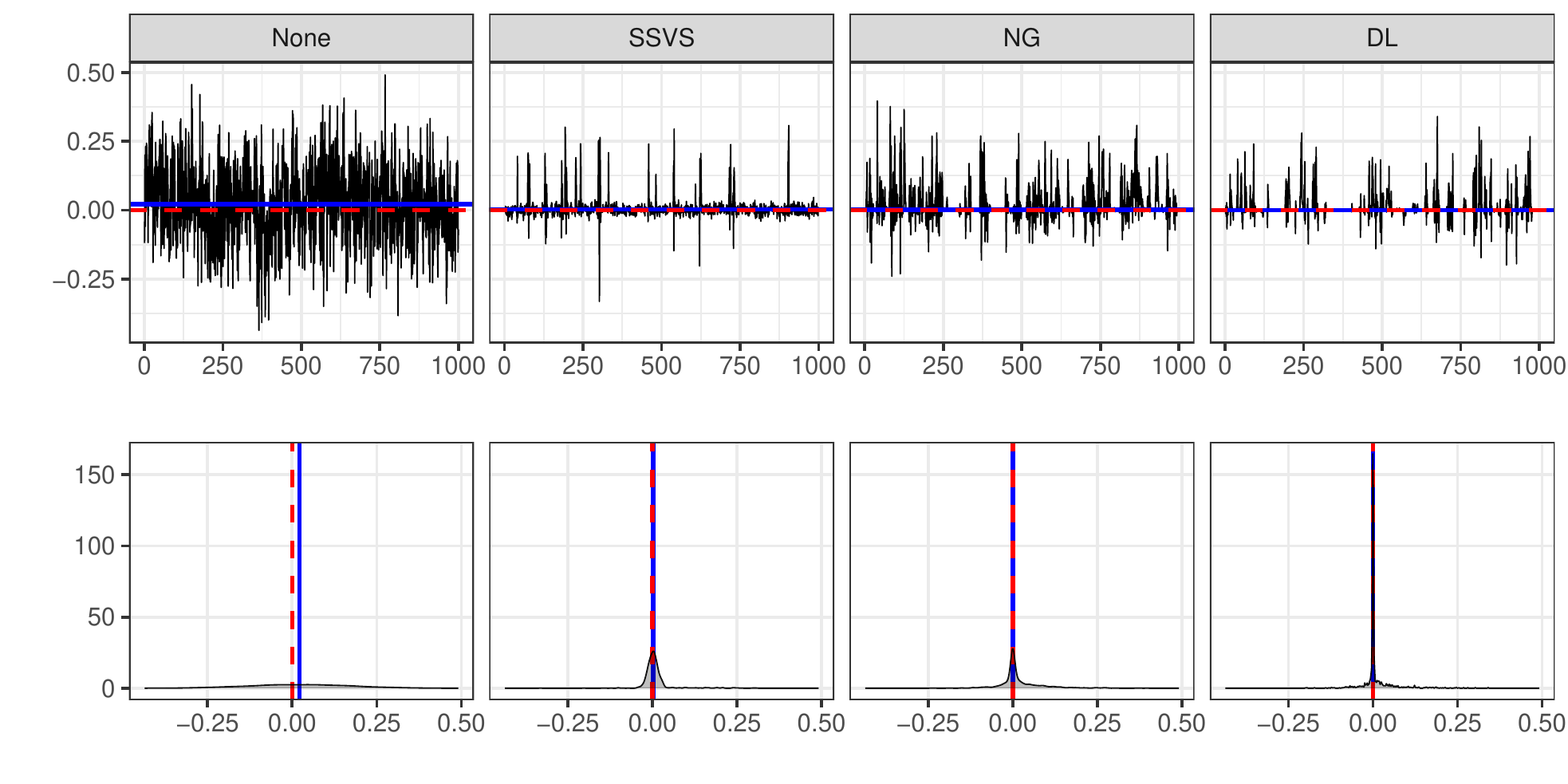}
  \caption{Diagnostic plots for a true coefficient equal to zero ($N=100$, $K = 50$ and $q = 10$).}\vspace*{-0.25cm}
  \caption*{\footnotesize{\textit{Notes:} The panel above shows trace plots for the specifications under scrutiny (horizontal axis: number of iteration; vertical axis: parameter value), while the lower panel depicts the posterior distribution of the parameter (horizontal axis: parameter value; vertical axis: density). The dashed red line indicates the true parameter value, the blue line is the median of the obtained posterior distribution.}}
  \label{plot:zero}
\end{figure*}
\begin{figure*}[!htbp]
  \includegraphics[width=\textwidth]{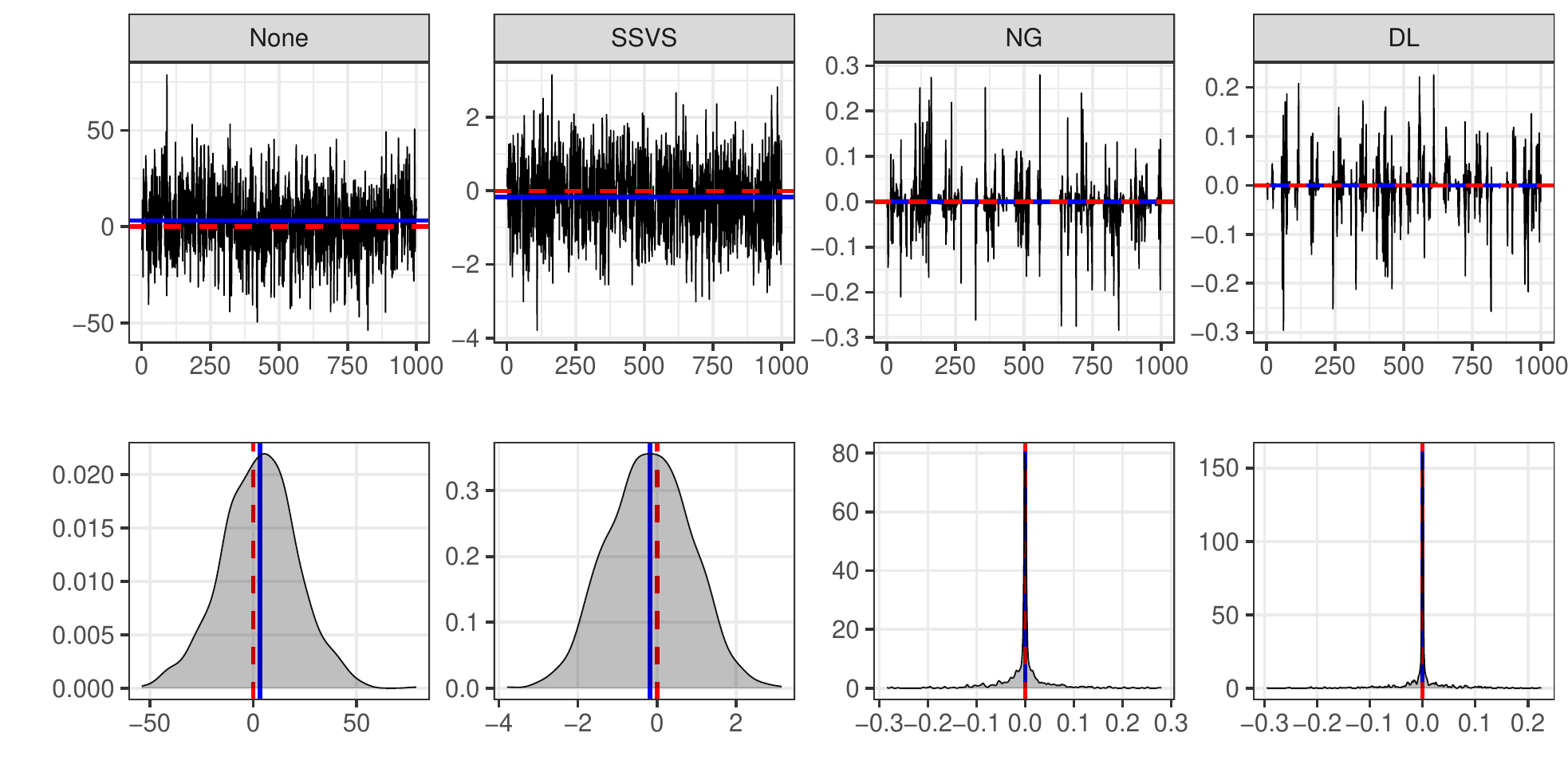}
  \caption{Diagnostic plots for all prior setups of a true coefficient equal to zero ($K = 150$ and $q = 10$).}\vspace*{-0.25cm}
  \caption*{\footnotesize{\textit{Notes:} The panel above shows trace plots for the specifications under scrutiny (horizontal axis: number of iteration; vertical axis: parameter value), while the lower panel depicts the posterior distribution of the parameter (horizontal axis: parameter value; vertical axis: density). The dashed red line indicates the true parameter value, the blue line is the median of the obtained posterior distribution.}}
  \label{plot:zero300}
\end{figure*}

The first example, for $K = 50$ exogenous covariates, is presented in \autoref{plot:zero}. The upper panel shows trace plots for all prior specifications, while the lower depicts the marginal posterior density of the parameter. The dashed red line indicates the true parameter value, the blue line is the median of the obtained posterior distribution. Even in this case, where ten out of the 50 slope coefficients in the parameter vector $\bm{\beta}$ are different from zero, we find that the standard specification without shrinkage (\textit{None}) of the parameter space results in a comparatively high variance of parameter estimates, observable in \autoref{plot:zero}. Considering the three shrinkage approaches, we find that the DL and NG priors concentrate more point mass on zero by design, while the SSVS prior shows considerable variation in the interval around zero.

A more interesting case is given in \autoref{plot:zero300}. In this setup we include $150$ covariates, where only ten exhibit non-zero values by construction. As is evident considering \textit{None}, the variance around parameter estimates is huge when the number of parameters to estimate exceeds the number of observations. Since the semi-automatic setup of the SSVS prior set forth in \cref{subsec:ssvs} relies on these quantities to scale prior hyperparameters, this is influential regarding resulting parameter estimates for the SSVS approach. Notice that the obtained scaling factor renders the variance of the spike component comparatively large due to the specific setup involved, and SSVS does not provide enough shrinkage, as shown in \autoref{plot:zero300}. 

This issue is not observable in the case of the DL and NG priors, providing evidence for their excellent adaptive shrinkage properties in high-dimensional settings. For the NG prior, it is worth mentioning that the point mass placed on zero is comparable to the case for $K = 50$. By the fact that the single hyperparameter of the DL prior is set to $1/K$ in the default case, higher dimensional problems result in even stronger shrinkage towards zero for true zero coefficients. Evidence of the increased tightness of the prior can be observed with regard to the density scales in \autoref{plot:zero300}. Moreover, note that we also observe that the shrinkage priors are able to recover coefficients close to zero equally well. 

\begin{figure*}[t]
  \includegraphics[width=\textwidth]{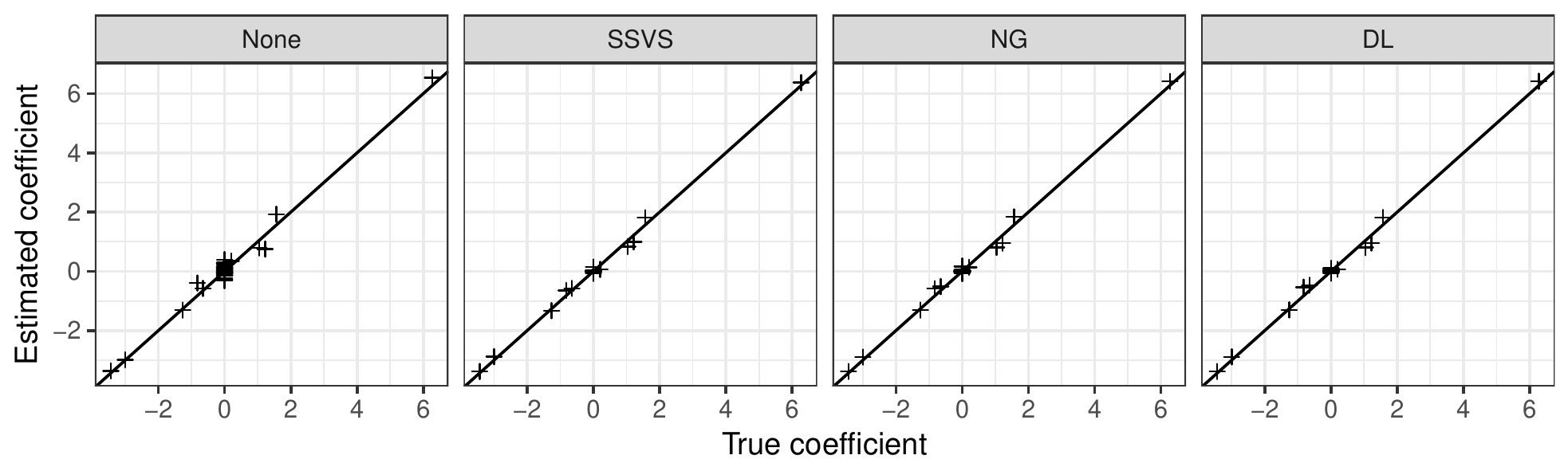}
  \caption{Comparison of posterior estimates to true parameter values for $K = 50$ and $q = 10$.}\vspace*{-0.25cm}
  \caption*{\footnotesize{\textit{Notes:} The panel above shows the correlation between true parameter values and estimated quantities based on the mean of the posterior draws for scenario 1. The black line indicates the 45 degree line.}}
  \label{plot:comparison1}
\end{figure*}
\begin{figure*}[t]
  \includegraphics[width=\textwidth]{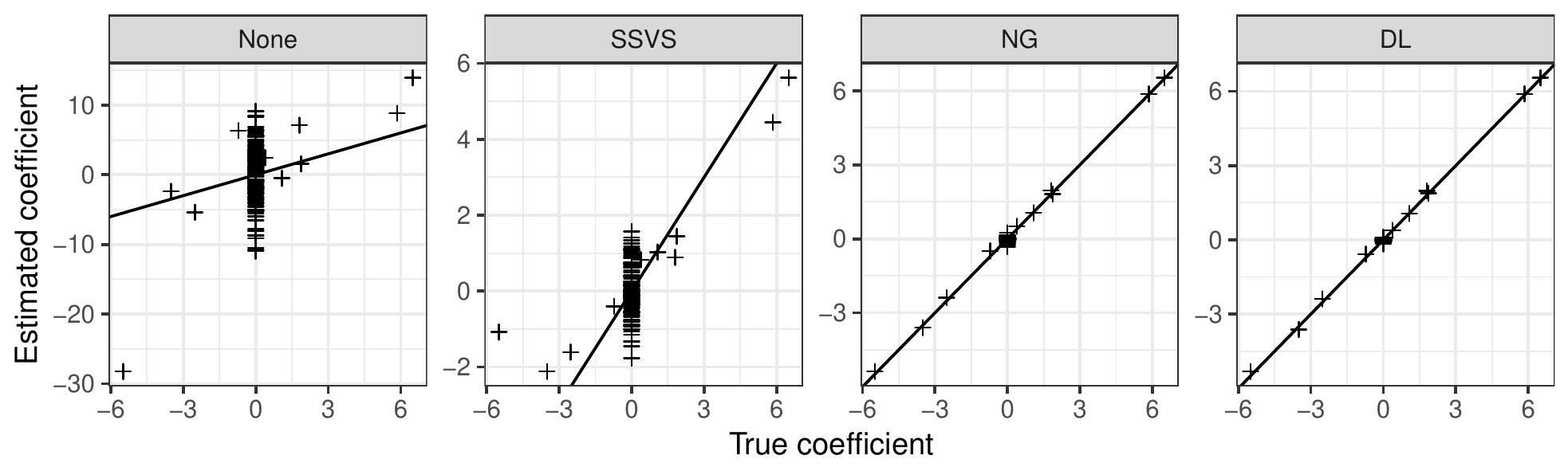}
  \caption{Comparison of posterior estimates to true parameter values for $K = 150$ and $q = 10$.}\vspace*{-0.25cm}
  \caption*{\footnotesize{\textit{Notes:} The panel above shows the correlation between true parameter values and estimated quantities based on the mean of the posterior draws. The black line indicates the 45 degree line.}}
  \label{plot:comparison101}
\end{figure*}

Following this brief description of diagnostic plots, we again use the scenarios above as examples and consider all of the resulting parameter estimates jointly. Plots of the true regression coefficients against their posterior medians are depicted in \autoref{plot:comparison1} and \autoref{plot:comparison101}. The black line indicates the 45 degree line, reflecting the objective of perfectly estimated parameters. In the low-dimensional example given in \autoref{plot:comparison1}, we find that point estimates based on the median of obtained posterior distributions for the parameters are mostly correct also in the case where no shrinkage is applied. However, note that these estimates are rather imprecise, which renders them suboptimal in terms of significance for interpretation, and also for applications involving predictions. The three shrinkage approaches closely mirror each other, even though it is worth noting that the NG and DL prior allow for even more precise estimates in terms of the variance.

A completely different case emerges for $150$ covariates. \autoref{plot:comparison101} showcases severe problems regarding the standard prior specification without shrinkage (\textit{None}). Even though the SSVS prior performs slightly better than this approach, it is evident that the default setup regarding the scaling of the hyperparameters is suboptimal in high-dimensions. In particular, and as we will discuss below, the SSVS prior actually tracks non-zero coefficients quite well, but imprecisely. Issues arise mainly from its disability to capture zero coefficients adequately in the given amount of iterations of the MCMC algorithm, pointing towards mixing problems and suboptimal convergence as stated in \citet{doi:10.1080/01621459.2014.960967}. In this respect, both the NG and DL prior are superior. The extraordinary empirical properties of these priors are evident in \autoref{plot:comparison101}, where the majority of the coefficients is recovered almost perfectly. Note that the DL prior provides slightly tighter shrinkage of true zero coefficients.

\subsection{Evaluation of performance}\label{sec:results}
Turning to an overview of all estimated parameters of the models, with respect to a total of eight different scenarios given by combinations of $K = \{50,100,150,200\}$ and $q = \{10, 20\}$, we present root mean squared error (RMSE) measures.\footnote{We adopt the standard root mean squared error measure for the posterior median $\hat{\bm{\beta}}$, $\text{RMSE}(\hat{\bm{\beta}}) = (\sum_{k=1}^K(\hat{\beta}_k-\tilde{\beta}_k)^2/K)^{\frac{1}{2}}$ and also calculate the corresponding quantity with respect to each draw from the MCMC algorithm, $\text{RMSE}_\text{dr}(\hat{\bm{\beta}}) = \sum_{k=1}^K(\sum_{t=1}^{T}(\hat{\beta}_{k,t}-\tilde{\beta}_k)^2/T)^{\frac{1}{2}}/K$. This serves as a means to illustrate differing degrees of precision of the estimators.} The specific choice of the number of included covariates and number of non-zero coefficients implies that we have degrees of sparsity ranging from 5 to 40 percent for the parameter vector. Average results for $100$ iterations are presented in \autoref{tab:rmse-est} and \autoref{tab:rmse-dr}.

\begin{table*}[t]
\caption{Root mean squared error measures for point estimates.}\vspace*{-1.8em}
\footnotesize
\begin{center}
\begin{threeparttable}
\begin{tabular*}{\textwidth}{@{\extracolsep{\fill}} ll cccc cccc}
\toprule
& & \multicolumn{4}{c}{$q = 10$} & \multicolumn{4}{c}{$q = 20$}\\ 
\cmidrule(l{3pt}r{3pt}){3-6}\cmidrule(l{3pt}r{0pt}){7-10}
& $\text{RMSE}$ & $\bm{\beta}$ & $\sigma^2$ & $\rho$ & Time & $\bm{\beta}$ & $\sigma^2$ & $\rho$ & Time\\
\midrule
$K=50$    & None  & 0.0184  & 0.2319  & 0.0044  & 1.02    & 0.0180  & 0.3072  & 0.0028  & \textbf{1.00} \\ 
      & SSVS  & \textbf{0.0078}   & \textbf{0.0150}   & \textbf{0.0032}   & \textbf{1.00}   & \textbf{0.0107}   & 0.0189  & \textbf{0.0021}   & \textbf{1.00} \\ 
      & NG  & 0.0090  & 0.0155  & 0.0034  & 1.05    & 0.0111  & \textbf{0.0168}   & 0.0022  & 1.06 \\ 
      & DL  & 0.0089  & 0.0185  & 0.0035  & 1.01    & 0.0120  & 0.0311  & {0.0021}  & 1.02 \\
$K=100$   & None  & 0.4589  & $>10$   & 0.1606  & 1.02    & 0.4670  & $>10$   & 0.1191  & 1.04 \\ 
      & SSVS  & 0.0118  & 0.0398  & 0.0038  & 1.09    & 0.0130  & 0.0317  & 0.0025  & 1.06 \\ 
      & NG  & 0.0080  & 0.0297  & 0.0033  & 1.06    & \textbf{0.0101}   & \textbf{0.0296}   & \textbf{0.0021}   & 1.02 \\ 
      & DL  & \textbf{0.0073}   & \textbf{0.0182}   & \textbf{0.0032}   & \textbf{1.00}   & \textbf{0.0101}   & 0.0316  & 0.0023  & \textbf{1.00} \\
$K=150$   & None  & 0.3665  & $>10$   & 0.2914  & \textbf{1.00}   & 0.4847  & $>10$   & 0.2438  & \textbf{1.00} \\ 
      & SSVS  & 0.2578  & 0.0993  & 0.0304  & 1.11    & 0.1882  & 0.0973  & 0.0265  & 1.09 \\ 
      & NG  & 0.0075  & 0.0498  & \textbf{0.0035}   & 1.11    & 0.0088  & 0.0500  & 0.0026  & 1.09 \\ 
      & DL  & \textbf{0.0064}   & \textbf{0.0226}   & 0.0038  & 1.06    & \textbf{0.0080}   & \textbf{0.0304}   & \textbf{0.0023}   & 1.03 \\
$K=200$   & None  & 0.2885  & $>10$   & 0.2625  & \textbf{1.00}   & 0.3720  & $>10$   & 0.2522  & \textbf{1.00} \\ 
      & SSVS  & 0.1353  & 0.0958  & 0.0503  & 1.21    & 0.1354  & 0.1103  & 0.0367  & 1.18 \\ 
      & NG  & 0.0075  & 0.0789  & 0.0039  & 1.12    & 0.0087  & 0.0799  & 0.0027  & 1.09 \\ 
      & DL  & \textbf{0.0055}   & \textbf{0.0268}   & \textbf{0.0032}   & 1.02    & \textbf{0.0074}   & \textbf{0.0262}   & \textbf{0.0025}   & \textbf{1.00} \\ 
\bottomrule
\end{tabular*}
\begin{tablenotes}[para,flushleft]
\footnotesize{\textit{Notes}: $K$ denotes the number of covariates, $q$ is the number of non-zero parameters. Time refers to the relative time with respect to the fastest approach for the given scenario. RMSE is the root mean squared error.}
\end{tablenotes}
\end{threeparttable}
\end{center}
\label{tab:rmse-est}
\end{table*}

The results for the standard RMSE measure closely resemble the notions obtained in the discussion above. In particular, we find that the SSVS prior performs slightly better than other shrinkage approaches in low dimensional settings where $K < N$, both in terms of estimation errors for parameters in $\bm{\beta}$, $\sigma^2$, the spatial dependence parameter $\rho$ and also time elapsed for cycling through the MCMC algorithm. However, both the NG and DL prior show promising values regarding RMSEs and can be considered as feasible alternatives. For the case of $K = N$, we find a different picture. The approach without shrinkage (\textit{None}) performs rather poorly, evidenced by large errors in terms of all parameters. In particular, based on poor estimates for the coefficients in $\bm{\beta}$, this results in the inability to correctly estimate $\sigma^2$. We find particularly good performance measures for the DL prior with the NG prior being close second. This finding is particularly pronounced in the denser case with respect to the coefficient vector ($q=20$), where the NG prior outperforms the DL prior in terms of precision regarding the error variances and the spatial dependence parameter.

Turning to the higher dimensional cases where $K$ exceeds $N$ it is worth noting that SSVS runs into similar problems than the standard approach, where poor parameter estimates result in huge values for the error variances. Since this problem only occurred in a minor subset of the $100$ iterations per scenario, we chose to exclude these erroneous simulations. In this setup, the approach without shrinkage again appeared to perform poorly, as already observed in the case with $K = 100$. Second, due to the notion that SSVS typically shows suboptimal mixing and convergence properties in high dimensional settings -- stemming from the necessity of stochastic search over an enormous parameter space -- it appears that the default setting of $2,000$ iterations for the MCMC algorithm is not sufficient to obtain reasonable posterior distributions for the parameters. Finally, we find that the DL prior performs best regarding coefficients and error variances -- compared to the NG prior, which is only superior in terms of estimating the spatial dependence parameter $\rho$. Note that estimates in the case of more dense coefficient vectors ($q=20$) are typically slightly worse, due to the necessity of estimating more none-zero parameters with the same number of observations. A similar picture is present in the case of $K \gg N$. Mirroring the results already obtained in the previous scenario, the DL prior appears to be the fastest of the shrinkage priors. Interestingly, the superiority of the DL prior regarding estimates of the error variances gets even larger when compared to the worse estimates resulting from the NG prior. 

\begin{table*}[t]
\caption{Root mean squared error measures for density estimates.}\vspace*{-1.8em}
\footnotesize
\begin{center}
\begin{threeparttable}
\begin{tabular*}{\textwidth}{@{\extracolsep{\fill}} ll cccc cccc}
\toprule
& & \multicolumn{4}{c}{$q = 10$} & \multicolumn{4}{c}{$q = 20$}\\ 
\cmidrule(l{3pt}r{3pt}){3-6}\cmidrule(l{3pt}r{0pt}){7-10}
& $\text{RMSE}_\text{dr}$ & $\bm{\beta}$ & $\sigma^2$ & $\rho$ & Time & $\bm{\beta}$ & $\sigma^2$ & $\rho$ & Time\\
\midrule
$K=50$  & None      & 0.0250    & 0.0371    & 0.0072    & 1.02      & 0.0248    & 0.0363    & 0.0043    & {1.00} \\ 
    & SSVS      & \textbf{0.0096}     & \textbf{0.0273}     & \textbf{0.0051}     & \textbf{1.00}   & \textbf{0.0127}     & 0.0318    & \textbf{0.0034}     & \textbf{1.00} \\ 
    & NG      & 0.0122    & 0.0276    & 0.0053    & 1.05      & 0.0149    & \textbf{0.0300}     & \textbf{0.0034}     & 1.06 \\ 
    & DL      & 0.0130    & 0.0357    & 0.0058    & 1.01      & 0.0174    & 0.0513    & 0.0039    & 1.02 \\

$K=100$ & None      & 0.6730    & $>10$     & 0.2057    & 1.02      & 0.6795    & $>10$     & 0.1482    & 1.04 \\ 
    & SSVS      & 0.0171    & 0.0480    & 0.0058    & 1.09      & 0.0185    & 0.0452    & 0.0040    & 1.06 \\ 
    & NG      & \textbf{0.0113}     & 0.0381    & \textbf{0.0052}     & 1.06      & \textbf{0.0136}     & \textbf{0.0398}     & \textbf{0.0034}     & 1.02 \\ 
    & DL      & 0.0118    & \textbf{0.0367}     & 0.0056    & \textbf{1.00}     & 0.0154    & 0.0569    & 0.0041    & \textbf{1.00} \\ 

$K=150$ & None      & 2.1817    & $>10$     & 0.3142    & \textbf{1.00}     & 2.1749    & $>10$     & 0.2807    & \textbf{1.00} \\ 
    & SSVS      & 0.1812    & 0.1034    & 0.0314    & 1.11      & 0.1376    & 0.1141    & 0.0276    & 1.09 \\ 
    & NG      & \textbf{0.0106}     & 0.0549    & \textbf{0.0053}     & 1.11      & \textbf{0.0122}     & \textbf{0.0553}     & \textbf{0.0036}     & 1.09 \\ 
    & DL      & 0.0111    & \textbf{0.0396}     & 0.0061    & 1.06      & 0.0138    & 0.0596    & 0.0042    & 1.03 \\

$K=200$ & None      & 2.4875    & $>10$     & 0.3122    & \textbf{1.00}     & 2.3582    & $>10$     & 0.3091    & \textbf{1.00} \\ 
    & SSVS      & 0.0954    & 0.1424    & 0.0516    & 1.21      & 0.1106    & 0.1806    & 0.0377    & 1.18 \\ 
    & NG      & \textbf{0.0103}     & 0.0812    & \textbf{0.0051}     & 1.12      & \textbf{0.0117}     & 0.0822    & \textbf{0.0036}     & 1.09 \\ 
    & DL      & 0.0111    & \textbf{0.0422}     & 0.0053    & 1.02      & 0.0138    & \textbf{0.0631}     & 0.0044    & \textbf{1.00} \\ 
\bottomrule
\end{tabular*}
\begin{tablenotes}[para,flushleft]
\footnotesize{\textit{Notes}: $K$ denotes the number of covariates, $q$ is the number of non-zero parameters. Time refers to the relative time with respect to the fastest approach for the given scenario. $\text{RMSE}_\text{dr}$ is the root mean squared error based on all MCMC draws providing insights in the performance in terms of parameter precision.}
\end{tablenotes}
\end{threeparttable}
\end{center}
\label{tab:rmse-dr}
\end{table*}

The main results discussed above also hold in the case of the adapted RMSE focusing on density predictions in \autoref{tab:rmse-dr}. Recall that we use this measure for the purpose of gaining insight into the precision of the obtained parameter estimates. Interestingly, even though being inferior to the DL prior in terms of point estimates, we find that the NG prior produces more precise estimates around the true values of the parameters in the coefficient vector $\bm{\beta}$. However, this finding does not carry over to estimates of the error variances $\sigma^2$, where the DL prior outperforms the NG approach. The spatial dependence parameter is estimated precisely in all cases regarding sparsity and number of covariates within the NG and DL shrinkage prior framework, and in scenarios where the SSVS prior and the standard approach still perform well. 

\section{Empirical illustration}
\label{sec:application}
In this section we aim at illustrating the performance of the proposed model specification using real data on pan-European regional economic growth and its empirical determinants. Specifically, we consider a cross-regional spatial Durbin model specification (see, for example, \citealt{lesage-pace_introductionspatial}) which also allows for spatially lagged explanatory variables as potential covariates. The specification used can be written as
\begin{equation}
\label{eq:growth}
\boldsymbol{S}(\rho)\boldsymbol{y}=\boldsymbol{\iota_N} \alpha + \boldsymbol{X\beta} + \boldsymbol{WX\vartheta} + \bm{u},\quad \bm{u} \sim \text{N}(\bm{0},\sigma^2\bm{I}_N)
\end{equation}
where $\boldsymbol{y}$ denotes an $N$-dimensional column vector of regional economic growth rates of per capita gross value added and the $N\times 1$ error term $\bm{u}$ is defined as before as iid normal. The $N\times K$ matrix $\boldsymbol{X}$ contains the set of potential predictors. $\boldsymbol{W}$ is an $N\times N$ row-stochastic spatial weight matrix as defined before. $\boldsymbol{S}(\rho)$ denotes a matrix exponential spatial filter with corresponding scalar parameter $\rho$ as defined before. $\boldsymbol{WX}$ is the spatial lag of the explanatory variables and explicitly incorporates spatially lagged information of $\boldsymbol{X}$ from neighboring regions, resulting in the spatial Durbin model specification mentioned above. In this empirical illustration we follow the typical structure of (spatial) growth regressions by assuming that information in matrix $\boldsymbol{X}$ is measured at the beginning of the sample period (which is the year $2000$).



\subsection{Regions, spatial weights and data}
For the empirical illustration we use data on regional economic growth on a sample of $273$ European NUTS-2 regions of 28 European countries. The dependent variable in the regression framework is the average annual growth rate of per capita gross value added in the period $2001-2010$. A detailed list of the regions used in the application is given in \autoref{tab:regions} in \cref{app:dataapp}. Table \ref{tab:variables} provides detailed information on the set of potential covariates in the matrix exponential spatial growth specification.
 
The set of predictors is in line with recent empirical applications on regional economic growth. Specifically, the matrix of explanatory variables contains information on the initial level of economic growth rates, human capital endowments, proxies for knowledge capital stocks, regional population structure, infrastructure, the region-specific industry mix, and other socio-economic quantities. It is worth noting that the spatial lag of the (non-constant) explanatory variables in the spatial Durbin framework sketched above results in doubling the set of potential covariates depicted in \autoref{tab:variables}. For the specification of the spatial weight matrix $\boldsymbol{W}$ we used a $10$-nearest row-stochastic specification.\footnote{Robustness checks using alternative numbers of nearest neighbors appeared to exert negligible effects on the results.}

\subsection{Empirical results}
In this subsection we present the results of the stochastic search variable selection (SSVS) prior, along with Normal-Gamma (NG), and Dirichlet-Laplace (DL) shrinkage setups including alternative prior hyperparameter specifications. Specifically, for the Normal-Gamma shrinkage prior a natural candidate is the Bayesian LASSO \citep{doi:10.1198/016214508000000337}, achieved by setting the prior hyperparameter $\theta=1$. Alternatively, we also consider $\theta = 0.1$, which refers to the generalized version characterized by heavier overall shrinkage. For the Dirichlet-Laplace (DL) shrinkage prior, we consider $a = 1/2$, which presents a usual benchmark in empirical research, while $a = 1/K$ is the default prior specification based on theoretical considerations \citep[see][]{doi:10.1080/01621459.2014.960967}.

Table \ref{tab:results} depicts estimation results for the competing prior setups under scrutiny. Note that only covariates where the corresponding slope coefficient is statistically significant from zero based on the lower $10$ percent and upper $90$ posterior interval in at least one of the candidate specifications are reported. For each prior setup, \autoref{tab:results} reports the posterior means (labeled Mean) and corresponding posterior standard deviations (labeled SD) for the parameters under scrutiny. Both quantities are based on $3,000$ retained draws of the MCMC algorithms described in \cref{app:posteriors}.\footnote{Note that for all alternative specifications we used a number of $12,000$ iterations with the first $3,000$ serving as burn-ins. To reduce the autocorrelation of the draws for $\rho$ we have used \textit{thinning} by considering only every third draw (see, for example, \citealt{koop2003}), resulting in a total of $3,000$ draws for posterior inference.}

Highly significant estimates for the spatial parameter $\rho$ result in all specifications considered, ranging from $-0.811$ to $-0.972$. Using the correspondence between the spatial parameter ($\xi$) in conventional spatial autoregressive frameworks and its matrix exponential spatial counterpart ($\rho$) stated by \citet{LESAGE2007190}, we find an implied spatial autoregressive parameter $\xi$ between $0.56$ and $0.62$. This degree of spatial dependence resembles the findings in recent studies on regional spatial economic growth in Europe \citep[see, for example,][]{doi:10.1080/00343404.2012.678824,doi:10.1080/17421770802353758}.

Turning attention to the (non-constant) potential growth determinants, \autoref{tab:results} shows four slope coefficients that are statistically significant in all prior setups. These variables are the initial level of income, the old-age dependency ratio, as well as both the share of low-educated employment and its corresponding spatial lag. Overall, \autoref{tab:results} shows rather similar results for both the magnitudes of the estimated effects as well as their significance. However, a notable exception is the Bayesian LASSO setup NG($\theta=1$), which highlights a markedly higher amount of significant slope parameters as compared to the competing specifications.

As expected, the initial income variable appears to negatively affect regional economic growth rates in all specifications, providing evidence for conditional convergence among the regions in the sample. However, the SSVS setting also shows significant posterior mean of the spatially lagged initial income variable, pointing towards the existence of positive growth spillovers emanating from the level of income of neighboring regions. This means that regions seem to benefit from the spatial proximity of rich regions in terms of accelerated growth rates. The old-age dependency ratio (measured in terms of the ratio of the number of people aged 65 and over to the working age population) appears to exhibit a negative impact on regional economic growth rates. Except for the Bayesian LASSO specification (NG ($\theta=1$)), the estimated coefficients are rather similar. 

Interestingly, the results presented in \autoref{tab:results} corroborate the findings of previous studies (see, for example, \citealt{GEAN:GEAN12057}, or \citealt{doi:10.1080/00343404.2012.678824}) by detecting the share of low educated working age population (Lower education workers) as being more robustly correlated to regional economic growth as compared to a measure of tertiary education attainment (higher education workers). As expected, lower education workers appear to exhibit a negative impact on regional growth in all specifications considered. However, it is worth noting that the spatial lag of this variable appears to exhibit a positive impact, indicating that positive effects on income growth to neighboring regions. Work by \cite{olejnik2008}, for example, argue that an increase in the lower educated labor force might result in positive growth spillovers, by assuming that such an increase might be primarily due to migration of workers between regions.


\begin{table*}[!t]
\caption{Estimation results.}\vspace*{-1.8em}
\footnotesize
\begin{center}
\begin{threeparttable}
\begin{tabular*}{\linewidth}{@{\extracolsep{\fill}} l rc rc rc rc rc}
\toprule
& \multicolumn{2}{c}{SSVS} & \multicolumn{2}{c}{NG ($\theta = 1$)} & \multicolumn{2}{c}{NG ($\theta = 0.1$)} & \multicolumn{2}{c}{DL ($a = 1/2$)} & \multicolumn{2}{c}{DL ($a = 1/K$)}\\ 
\cmidrule(l{3pt}r{3pt}){2-3}\cmidrule(l{3pt}r{0pt}){4-5}\cmidrule(l{3pt}r{0pt}){6-7}\cmidrule(l{3pt}r{0pt}){8-9}\cmidrule(l{3pt}r{0pt}){10-11}
Variable & Mean & SD & Mean & SD & Mean & SD & Mean & SD & Mean & SD \\ 
\midrule
Initial income$^{\ast}$ & \textbf{-0.857} & 0.195 & \textbf{-0.702} & 0.194 & \textbf{-0.734} & 0.241 & \textbf{-0.669} & 0.225 & \textbf{-0.702} & 0.209 \\ 
Population density & -0.036 & 0.062 & \textbf{-0.205} & 0.107 & -0.050 & 0.074 & -0.036 & 0.068 & -0.031 & 0.063 \\ 
Child dependency ratio & -0.017 & 0.025 & \textbf{-0.057} & 0.028 & -0.018 & 0.026 & -0.011 & 0.021 & -0.009 & 0.019 \\ 
Old-age dependency ratio$^{\ast}$ & \textbf{-0.038} & 0.022 & \textbf{-0.055} & 0.020 & \textbf{-0.036} & 0.022 & \textbf{-0.028} & 0.022 & \textbf{-0.027} & 0.022 \\ 
Lower education workers$^{\ast}$ & \textbf{-0.060} & 0.009 & \textbf{-0.060} & 0.011 & \textbf{-0.056} & 0.012 & \textbf{-0.048} & 0.016 & \textbf{-0.051} & 0.014 \\ 
$\bm{W}$ Initial income & \textbf{0.569} & 0.388 & 0.173 & 0.248 & 0.251 & 0.294 & 0.140 & 0.242 & 0.178 & 0.270 \\ 
$\bm{W}$ Child dependency ratio & 0.021 & 0.033 & \textbf{0.080} & 0.054 & 0.020 & 0.035 & 0.015 & 0.030 & 0.012 & 0.027 \\ 
$\bm{W}$ Agrilculture employment & 0.007 & 0.012 & \textbf{0.033} & 0.025 & 0.002 & 0.010 & 0.002 & 0.011 & 0.001 & 0.009 \\ 
$\bm{W}$ Construction employment & 0.019 & 0.037 & \textbf{0.123} & 0.088 & 0.013 & 0.037 & 0.018 & 0.042 & 0.014 & 0.037 \\ 
$\bm{W}$ Technology resources & 0.004 & 0.012 & \textbf{0.054} & 0.035 & 0.003 & 0.011 & 0.002 & 0.013 & 0.002 & 0.010 \\ 
$\bm{W}$ Higher education workers & -0.003 & 0.012 & \textbf{-0.050} & 0.032 & -0.002 & 0.010 & -0.003 & 0.013 & -0.001 & 0.009 \\ 
$\bm{W}$ Lower education workers$^{\ast}$ & \textbf{0.048} & 0.013 & \textbf{0.055} & 0.015 & \textbf{0.043} & 0.015 & \textbf{0.030} & 0.020 & \textbf{0.033} & 0.019 \\ 
\midrule
  $\sigma^2$ & \textbf{0.642} & 0.057 & \textbf{0.637} & 0.060 & \textbf{0.657} & 0.081 & \textbf{0.671} & 0.065 & \textbf{0.667} & 0.064 \\ 
  $\rho$ & \textbf{-0.972} & 0.151 & \textbf{-0.811} & 0.146 & \textbf{-0.925} & 0.138 & \textbf{-0.848} & 0.134 & \textbf{-0.874} & 0.146 \\
\bottomrule
\end{tabular*}
\begin{tablenotes}[para,flushleft]
\footnotesize{\textit{Notes}: The table contains a subset of all variables for which coefficients are significant in at least one specification ($^{\ast}$significant in all specifications). Estimates depicted in bold are statistically different from zero based on the 80 percent posterior credible interval.}
\end{tablenotes}
\end{threeparttable}
\end{center}
\label{tab:results}
\end{table*}

\section{Concluding remarks}
\label{sec:conclusions}
Dealing with model uncertainty in spatial autoregressive model specifications has been subject to numerous studies in general, especially in the empirical economic growth literature. However, spatial econometric applications typically rely on Bayesian model-averaging techniques, which suffer from severe drawbacks both in terms of computational time and possible extensions to more flexible model specifications. In spatial autoregressive models, the computational burden emanates from the calculation of marginal likelihoods, where no closed form solutions are available. Recent contributions to the literature as a means to alleviating the computational difficulties include a variant of the conventional stochastic search variable selection prior discussed in \citet{doi:10.1080/17421772.2016.1227468}. However, shortcomings of this approach include slow mixing and convergence properties in the presence of a large number of available covariates and difficulties in choosing prior hyperparameters.

This paper aims at generalizing two absolutely continuous hierarchical shrinkage priors -- the Normal-Gamma and the Dirichlet-Laplace shrinkage prior -- to the matrix exponential spatial specification in order to alleviate the above-mentioned drawbacks of both Bayesian model averaging and standard stochastic search variable selection priors. The proposed frameworks allow for flexible and adaptive, but also computationally efficient stochastic variable selection, where extensions to basic spatial model specification can be easily implemented in Markov chain Monte Carlo sampling algorithms. For illustrative purposes, and to evaluate prior-specific properties in the presence of spatial dependence, the paper carries out an extensive simulation study using synthetic data sets. An empirical illustration is given by a study on economic growth of European regions.

Our results indicate that standard stochastic search variable selection priors work particularly well in relatively low-dimensional settings. However, severe problems occur in high-dimensional settings, where the number of potential covariates relative to the number of available observations becomes large. The proposed global-local shrinkage priors provide the required adaptiveness of shrinkage in a flexible and computationally efficient way. The Normal-Gamma shrinkage prior excels in terms of precision of estimates, while the Dirichlet-Laplace shrinkage prior slightly outperforms the former in terms of point estimations of parameters. Both proposed shrinkage priors can be considered valuable tools as a means to incorporate model uncertainty in spatial autoregressive frameworks, regardless of dimensionality of the problem at hand.

\small\singlespacing
\bibliographystyle{cit_econometrica}
\bibliography{lit}
\addcontentsline{toc}{section}{References}

\normalsize\onehalfspacing\newpage
\begin{appendices}\crefalias{section}{appsec}
\setcounter{equation}{0}
\renewcommand\theequation{A.\arabic{equation}}

\section{Stochastic Search Variable Selection}\label{subsec:ssvs}
A standard variant of the SSVS prior, similar to the one employed in \cite{Piribauer2016} and \citet{doi:10.1080/17421772.2016.1227468} for conventional spatial autoregressive models assumes each element of the parameter vector $\bm{\beta}$ to come from a mixture of two Gaussians,
\begin{equation}
\beta_r | \delta_r \sim \delta_r \text{N}(0,\ubar{s}_{1r}) + (1-\delta_r) \text{N}(0,\ubar{s}_{0r}),
\end{equation}
for $r = 1,\hdots,K$ with $\ubar{s}_{0r} \ll \ubar{s}_{1r}$ resulting in the spike-and-slab combination of components centered on zero. Variance hyperparameters are set for all $\beta_r$, that is $\bm{\ubar{s}}_j = (\ubar{s}_{j1},\hdots,\ubar{s}_{jK})'$ for $j\in\{0,1\}$. We use the semi-automatic approach to scale the hyperparameters in $\ubar{\bm{s}}_j$ by estimating the saturated model without applying shrinkage and obtain the posterior variance covariance matrix $\hat{\bm{\Sigma}}$. Subsequently, we set the spike component variances to $\bm{\ubar{s}}_0 = 0.01\ \text{diag}(\hat{\bm{\Sigma}})$, and the slab component to $\bm{\ubar{s}}_1 = 100\ \text{diag}(\hat{\bm{\Sigma}})$. 

Moreover, we introduce $\bm{\delta} = (\delta_1,\hdots,\delta_K)'$, which is a vector of latent quantities, binarily indicating which mixture component is active with respect to the $r$th coefficient $\beta_r$. This can be exploited to calculate posterior inclusion probabilities for all covariates, establishing a nexus to the model-averaging framework regarding interpretation. To estimate latent inclusion indicators, we hierarchically impose prior inclusion probabilities $\ubar{\omega}_r$, chosen by the researcher, on the latent parameters
\begin{equation}
\bm{\delta} = \prod_{r=1}^{K} \ubar{\omega}_r^{\delta_r} (1-\ubar{\omega}_r)^{1-\delta_r}.
\end{equation}
\citet{10.2307/24306083} provide an intuitive interpretation for $p(\delta_r=1) = 1 - p(\delta_r=0) = \ubar{\omega}_r$, which captures whether $\beta_r$ is large enough to justify including the corresponding explanatory variable in the model. Arguably, a natural choice arising for $\omega_r~(r = 1,\hdots,K)$ is $0.5$. The conditional posterior of $\bm{\delta}$ is a Bernoulli distribution, as presented by \citet{doi:10.1080/01621459.1993.10476353,10.2307/24306083},
\begin{align}\label{eq:ssvs-delta}
p(\delta_r=1|\bullet) &\sim \text{Ber}\left(\obar{u}_{1r}/(\obar{u}_{0r}+\obar{u}_{1r})\right),\\
\obar{u}_{1r} &= \ubar{s}_{1r}^{-1}\exp\left(-\beta_r^2/2\ubar{s}_{1r}^2\right)\ubar{\omega}_r,\nonumber\\
\obar{u}_{0r} &= \ubar{s}_{0r}^{-1}\exp\left(-\beta_r^2/2\ubar{s}_{0r}^2\right)\left(1-\ubar{\omega}_r\right).\nonumber
\end{align}
The coefficient vector $\bm{\beta}$ is sampled using \autoref{eq:theta-ssvs}. Elements on the main diagonal of the prior variance-covariance matrix $\ubar{\bm{\Sigma}}$ whose latent indicator in $\bm{\delta}$ is equal to one are set to $\ubar{s}_{1r}$, implying less prior influence and allowing for unrestricted variation. In the opposite case $\ubar{s}_{0r}$ is set, resulting in shrinkage towards zero.

\newpage\section{Bayesian MCMC estimation}\label{app:posteriors}
Estimation is carried out running an Markov-chain Monte Carlo (MCMC) algorithm. One picks starting values for all parameters of the model and repeats the steps described below for a sufficient amount of iterations. The researcher decides on the number of iterations that depends on the mixing and convergence properties of the data at hand, and discards a number of initial draws as burn-in. Specifically, the sampler cycles through the following routine:
\begin{enumerate}[label=(\roman*),wide=0pt,leftmargin=*]
  \item Given the values for parameters from the most recent iteration, draw $\sigma^2$ from the inverse Gamma distribution given in \autoref{eq:sigma2}.
  \item This step depends on the specific shrinkage prior employed, items below are mutually exclusive:
  \begin{itemize}[leftmargin=*]
    \item \textbf{SSVS prior}: Draw $\bm{\delta}$ from the Bernoulli distributed conditional posterior in \autoref{eq:ssvs-delta}, then update the prior variance-covariance matrix as described in the text and use it to sample from the Gaussian in \autoref{eq:theta-ssvs}.
    \item \textbf{NG prior}: The local parameters $\phi_r$ are sampled independently from a generalized inverse Gaussian distribution\footnote{We use the algorithm described and implemented by \citet{Hoermann2014} to sample from a generalized inverse Gaussian distribution.} as in \autoref{eq:ng-psi}, while the global shrinkage parameter $\lambda$ comes from a Gamma distribution depicted in \autoref{eq:ng-lambda}. Again, update the variance-covariance matrix $\text{diag}(\ubar{\bm{\Sigma}}) = \bm{\psi}$ and sample the parameters using \autoref{eq:theta-ssvs}.
    \item \textbf{DL prior}: Local and global scaling parameters are sampled from generalized inverse Gaussians given in \autoref{eq:dl-varphi} and \autoref{eq:dl-tau}, respectively. Draw the auxiliary variables $T_r~(1,\hdots,K)$ mentioned above again from a generalized inverse Gaussian and set $\phi_r = T_r/\sum_{j=1}^{K} T_j$. As in the case of the SSVS and NG prior, the variance-covariance matrix must be updated $\text{diag}(\ubar{\bm{\Sigma}}) = (\varphi_1 \phi_1^2 \tau^2, \hdots, \varphi_K \phi_K^2 \tau^2)'$ and used to sample from \autoref{eq:theta-ssvs}.
  \end{itemize}
  \item Update $\rho$ using a Metropolis-within-Gibbs step using the conditional posterior given in Section \ref{sec:shrinkage}.
\end{enumerate}

\newpage\section{Data Appendix}\label{app:dataapp}
\small\singlespacing
\renewcommand{\arraystretch}{1.5}
\begin{table*}[!ht]
\caption{Variables used in the empirical illustration}\vspace*{-1.8em}
\footnotesize
\begin{center}
\begin{threeparttable}
\begin{tabular*}{\textwidth}{@{\extracolsep{\fill}} p{0.25\linewidth} p{0.7\linewidth}}
\toprule
\textbf{Variable} & \textbf{Description}\\
\midrule
Gross fixed capital formation & physical capital measured in terms of gross fixed captital formation (per capita). \textit{Source}: Cambridge Econometrics \\ 
Initial income & Gross-value added divided by population, 2000. \textit{Source}: Cambridge Econometrics \\
Physical capital & Gross fixed capital formation, 2000. \textit{Source}: Cambridge Econometrics \\
Higher education workers & Share of population (aged 25 and over, 2000)  with higher education (ISCED levels 1-2). \textit{Source}: Eurostat \\
Lower education workers & Share of population (aged 25 and over, 2000) with lower education (ISCED levels 5-6). \textit{Source}: Eurostat \\

Technology resources & Human resources in science and technology, share in persons employed, 2000. \textit{Source}: Eurostat \\
Agricultural employment & Share of NACE A and B (agriculture) in total employment, 2000. \textit{Source}: Cambridge Econometrics \\
Manufacturing \& construction employment & Share of NACE C to F (mining, manufacturing, energy and construction) in total employment, 2000. \textit{Source}: Cambridge Econometrics\\
Construction employment & Share of NACE F (construction) in total employment, 2000.\\ 
Market services employment & Share of NACE G to K (market services) in total employment, 2000. \textit{Source}: Cambridge Econometrics\\
Output density & Gross-value added per square km, 2000. \textit{Source}: Eurostat\\
Employment density & Employed persons per square km, 2000. \textit{Source}: Eurostat\\
Population density & Population per square km, 2000. \textit{Source}: Eurostat\\
Population growth & Average growth rate of the population for 1996-2000. \textit{Source}: Eurostat\\
Labor force participation & Employed and unemployed persons as a share of total \\
Child dependency ratio & The ratio of the number of people aged 0-14 to the number of people aged 15-64, 2000. \textit{Source}: Eurostat\\
Old-age dependency ratio & The ratio of the number of people aged 65 and over to the number of people aged 15-64, 2000. \textit{Source}: Eurostat\\
Accessibility road & Potential accessibility road, ESPON space=100. \textit{Source}: ESPON\\
Accessibility rail & Potential accessibility rail, ESPON space=100. \textit{Source}: ESPON\\
\bottomrule
\end{tabular*}
\begin{tablenotes}[para,flushleft]
\end{tablenotes}
\end{threeparttable}
\end{center}
\label{tab:variables}
\end{table*}

\renewcommand{\arraystretch}{1}
\begin{table*}[!t]
\caption{European regions in the sample.}\vspace*{-0.8em}
\scriptsize
\begin{threeparttable}
\begin{tabular*}{\linewidth}{@{\extracolsep{\fill}} l p{0.82\linewidth}}
\toprule
\textbf{Country} & \textbf{Region} \\
\midrule
Austria	&	Burgenland, Salzburg, K\"arnten, Steiermark, Nieder\"osterreich, Tirol, Ober\"osterreich,	Vorarlberg, Wien	\\
Belgium	&	Prov. Antwerpen,	Prov. Luxembourg, Prov. Brabant Wallon, Prov. Namur,	Prov. Hainaut, Prov. Oost-Vlaanderen, Prov. Li\`ege, Prov. Vlaams Brabant, Prov. Limburg,	Prov. West-Vlaanderen, R\'egion de Bruxelles-Capitale	\\
Bulgaria &	Severen tsentralen, Yugoiztochen, Severoiztochen,	Yugozapaden, Severozapaden,	Yuzhen tsentralen	\\
Czech Republic	&	Jihov\'ychod,	Severoz\'apad, Jihoz\'apad, Stredn\'i Cechy, Moravskoslezsko,	Stredn\'e Morava,	Praha,	Severov\'ychod\\
Denmark	&	Hovedstaden, Sjaelland, Midjylland, Syddanmark, Nordjylland \\
Estonia	&	Estonia	\\
Finland	&	\AA land,	L\"ansi-Suomi, Etel\"a-Suomi, Pohjois-Suomi, It\"a-Suomi	\\
France	&	Alsace,	$\hat{\mbox{I}}$le de France, Aquitaine, Languedoc-Roussillon, Auvergne,	Limousin, Basse-Normandie,	Lorraine, Bourgogne,	Midi-Pyr\'en\'ees, Bretagne,	Nord - Pas-de-Calais,Centre,	Pays de la Loire, Champagne-Ardenne,	Picardie, Corse,	Poitou-Charentes, Franche-Comt\'e, Provence-Alpes-C$\hat{\mbox{o}}$te d'Azur, Haute-Normandie, Rh$\hat{\mbox{o}}$ne-Alpes	\\
Germany	&	Arnsberg,	Leipzig, Berlin,	Mecklenburg-Vorpommern, Brandenburg,	Mittelfranken, Braunschweig,	M\"unster, Bremen,	Niederbayern, Chemnitz,	Oberbayern, Darmstadt,	Oberfranken, Detmold, Oberpfalz, Dresden,	Rheinhessen-Pfalz, D\"usseldorf, Saarland, Freiburg, Sachsen-Anhalt, Giessen, Schleswig-Holstein, Hamburg, Schwaben, Hannover, Stuttgart, Karlsruhe, Th\"uringen, Kassel, Trier, Koblenz, T\"ubingen, K\"oln, Unterfranken, L\"uneburg, Weser-Ems	\\
Greece	&	Anatoliki Makedonia, Thraki, Kriti, Attiki,	Notio Aigaio, Dytiki Ellada, Peloponnisos, Dytiki Makedonia,	Sterea Ellada, Ionia Nisia,	Thessalia, Ipeiros, Voreio Aigaio, Kentriki Makedonia	\\
Hungary	&	D\'el-Alf\"old, K\"oz\'ep-Dun\'ant\'ul, D\'el-Dun\'ant\'ul, K\"oz\'ep-Magyarorsz\'ag, \'Eszak-Alf\"old, Nyugat-Dun\'ant\'ul, \'Eszak-Magyarorsz\'ag	\\
Ireland	&	Border, Midlands and Western, Southern and Eastern\\
Italy	&	Abruzzo,	Molise, Basilicata, Piemonte, Calabria, Bolzano-Bozen, Campania, Trento, Emilia-Romagna,	Puglia, Friuli-Venezia Giulia,	Sardegna, Lazio,	Sicilia, Liguria, Toscana, Lombardia,	Umbria, Marche,	Valle d'Aosta, Veneto	\\
Latvia	&	Latvia	\\
Lithuania	&	Lithuania	\\
Luxembourg	&	Luxembourg (Grand-Duch\'e)	\\
Netherlands	&	Drenthe, Noord-Brabant, Flevoland,	Noord-Holland, Friesland,	Overijssel, Gelderland, Utrecht, Groningen, Zeeland, Limburg, Zuid-Holland	\\
Norway & Agder og Rogaland, Sor-Ostlandet, Hedmark og Oppland, Trondelag, Nord-Norge, Vestlandet, Oslo og Akershus \\
Poland	&	Dolnoslaskie, Podkarpackie, Kujawsko-Pomorskie,	Podlaskie, Lodzkie, Pomorskie, Lubelskie, Slaskie, Lubuskie,	Swietokrzyskie, Malopolskie, Warminsko-Mazurskie, Mazowieckie,	Wielkopolskie, Opolskie,	Zachodniopomorskie	\\
Portugal	&	Alentejo, Lisboa, Algarve, Norte, Centro\\
Romania	&	Bucuresti-Ilfov, Sud-Muntenia, Centru, Sud-Est, Nord-Est, Sud-Vest Oltenia, Nord-Vest, Vest	\\
Slovak Republic	&	Bratislavsk\'y kraj,	V\'ychodn\'e Slovensko,	Stredn\'e Slovensko,	Z\'apadn\'e Slovensko	\\
Slovenia	&	Vzhodna Slovenija, Zahodna Slovenija	\\
Spain	&	Andalucia, Extremadura, Arag\'on, Galicia, Cantabria, Illes Balears, Castilla y Le\'on, La Rioja, Castilla-la Mancha, Pais Vasco, Catalu\~na, Principado de Asturias, Comunidad de Madrid, Regi\'on de Murcia, Comunidad Foral de Navarra, Comunidad Valenciana	\\
Sweden	&	Mellersta Norrland, Sm\aa land med \"oarna, Norra Mellansverige, Stockholm, \"Ostra Mellansverige, Sydsverige, \"Ovre Norrland, V\"astsverige	\\
Switzerland & Central Switzerland, Northwestern Switzerland, Eastern Switzerland, Ticino, Espace Mittelland, Zurich, Lake Geneva\\
United Kingdom	&	Bedfordshire, Hertfordshire,	Kent, Berkshire, Bucks and Oxfordshire,	Lancashire, Cheshire,	Leicestershire, Rutland and Northants, Cornwall and Isles of Scilly,	Lincolnshire, Cumbria,	Merseyside, Derbyshire and Nottinghamshire,	North Yorkshire, Devon,	Northern Ireland, Dorset and Somerset,	Northumberland, Tyne and Wear, East Anglia,	Outer London, East Riding and North Lincolnshire,	Shropshire and Staffordshire, East Wales,	South Western Scotland, Eastern Scotland,	South Yorkshire, Essex,	Surrey East and West Sussex, Gloucestershire Wiltshire,	Tees Valley and Durham, North Somerset, Greater Manchester,	West Midlands, Hampshire and Isle of Wight,	West Wales and The Valleys, Herefordshire, Worcestershire and Warks,	West Yorkshire, Inner London\\
\bottomrule
\end{tabular*}
\end{threeparttable}
\label{tab:regions}
\end{table*}

\end{appendices}

\end{document}